\documentclass[fleqn,usenatbib]{mnras}
\usepackage{newtxtext,newtxmath}
\usepackage[T1]{fontenc}
\DeclareRobustCommand{\VAN}[3]{#2}
\let\VANthebibliography\thebibliography
\def\thebibliography{\DeclareRobustCommand{\VAN}[3]{##3}\VANthebibliography}
\usepackage{graphicx}	
\usepackage{amsmath}	

\begin{document}

\title{Recent astrophysical observations reproduced by a short-range correlated van der Waals-type model?}

\author[E. H. Rodrigues, M. Dutra, and O. Louren\c{c}o]{E. H. Rodrigues$^1$, M. Dutra$^{1,2}$, and O. Louren\c{c}o$^{1,2}$
\\
$^1$Departamento de F\'isica, Instituto Tecnol\'ogico de Aeron\'autica, DCTA, 12228-900, S\~ao Jos\'e dos Campos, SP, Brazil
\\
\mbox{$^2$Universit\'e de Lyon, Universit\'e Claude Bernard Lyon 1, CNRS/IN2P3, IP2I Lyon, UMR 5822, F-69622, Villeurbanne, France} \\
}

\date{\today}

\maketitle

\begin{abstract}
We perform an improvement in a van der Waals-type model by including it effects of short-range correlations~(SRC). Attractive and repulsive parts of the nucleon-nucleon interaction are assumed to be density-dependent functions, more specifically, we adopt the Carnahan–Starling (CS) method for the latter, and a suitable expression for the former in order to reproduce the structure of the Clausius~(C) real gas model. The parametrizations of the resulting model, named as \mbox{CCS-SRC} model, are shown to be capable of reproducing the flow constraint at the high-density regime of symmetric nuclear matter for incompressibility values inside the range of $K_0=(240\pm 20)$~MeV. In the context of stellar matter, our findings point out a good agreement of the \mbox{CCS-SRC} model with recent astrophysical observational data, namely, mass-radius contours and dimensionless tidal deformability regions and values, coming from gravitational waves data related to the GW170817 and GW190425 events, and from the NASA’s Neutron star Interior Composition Explorer (NICER) mission. Furthermore, the values for the symmetry energy slope of the model ($L_0$) are in agreement with a recent range found for this quantity, claimed to be consistent with results reported by the updated lead radius experiment (PREX-2) collaboration. In this case, higher values of $L_0$ are favored, while the opposite scenario does not allow simultaneous compatibility between the model and the astrophysical data.
\end{abstract}


\pubyear{2023}

\section{Introduction} 

The universe is composed of many kinds of interesting structures, among them compact objects, namely, remnants of massive stars (mass bigger than six to eight times that of our sun), black holes, white dwarfs, and neutron stars (NSs). NSs are one of the densest existing objects and their internal structure is not fully known~\citep{debora-universe}. The number of NSs is estimated to be around one billion in Milky Way, but only a few thousand have been already observed~\citep{camenzind2007}. However, this scenario has been changing with advances in observational technologies of gravitational waves from high-frequency telescopes. Operating missions such as NASA's Neutron star Interior Composition Explorer Mission (NICER)~\citep{nicer}, and interferometers such as the Laser Interferometer Gravitational-Wave Observatory (LIGO)~\citep{LigoCol}, and the Virgo gravitational-wave detector~\citep{2012virgo}, hosted by the European Gravitational Observatory (EGO), have increased on a daily basis the number of detected objects. The fourth LIGO observation run will start in 2023 and projects a sensitivity goal of 160-190~Mpc for binary neutron star mergers, which means an increase in sensitivity of 35\%, and certainly, the emergence of a large amount of new data~\citep{calltech23}. 

The high-density environment found in NSs makes them excellent natural laboratories for the study and application of different models based on relativistic/nonrelativistic hadronic physics. Concerning the theory used to construct such models, it is worth mentioning that there are at least two conceptions of approaching. One of them uses the available nucleon-nucleon interactions to perform Brueckner-Hartree-Fock calculations~\citep{bhf}. The second approach is based on the fitting of some many-nucleon observables, additionally using the mean-field approximation, that allows the derivation of equations of state~(EoS) used to describe nuclear properties in finite (nuclei) and infinite systems (symmetric and asymmetric infinite nuclear matter) at zero and finite temperature regimes. Many studies are dedicated to implementing this method, in which nonrelativistic and relativistic hadronic models are equally applied.  A particular class of hadronic models used to describe nuclear matter through the second approach takes into account relativistic effects presented by nucleons in the medium. The first version of these models was proposed in~\cite{walecka} and is based on Quantum Field Theory, where a Lorentz invariant Lagrangian density is the starting point from which all thermodynamical quantities are derived.  

Recently, a new type of relativistic model has been used to describe nuclear systems, such as those found in NSs. It is based on the classical van der Waals~(vdW) EoS, but generalized for quantum systems also including relativity in its structure~\citep{vovchenko2015a,vovchenko2015b,vovchenko2017a,vovchenko2017b}. A modified version of this model was proposed in~\cite{lourenco2019} and ~\cite{jpgnosso} where the authors developed an approach in which both parts of the interaction, repulsive and attractive, are depending on the nuclear density. The model was verified to be consistent with some nuclear/stellar matter constraints. Here we proceed to improve such a model by including on it an important phenomenology observed in nuclear systems, namely, the short-range correlations~(SRC): effect exhibited in pairs of non-independent nucleons that emerge with high relative momentum in some nuclei such as $^{12}\rm C$, $^{27}\rm Al$, $^{56}\rm Fe$ and $^{208}\rm Pb$. The effect was observed after the collision of these nuclei with highly energetic incident particles~\citep{sciencesrc1,naturesrc2,naturescr3,naturesrc4,hen2017,duer2019} in experiments performed, for instance, at the Thomas Jefferson National Accelerator Facility~\citep{science-src}. We investigate how SRC impact the excluded volume model used here, and show that the improved model is in agreement with the flow constraint established in~\cite{danielewicz2002}, and also with recent astrophysical observational data provided by LIGO and Virgo Collaboration, and by the analysis of the NICER mission observations. For the comparison with these specific data, we present the results of the model concerning the mass-radius profiles and tidal deformability related to the binary neutron stars system.

Our paper is organized as follows: in Sec.~\ref{ddvdw} we present the main features of the density-dependent vdW model constructed in~\cite{lourenco2019,jpgnosso} on which our improved model is based on. Then, in Sec.~\ref{src} we develop the inclusion of SRC and show how such effects change the previous model for symmetric nuclear matter and stellar matter. For both cases, we show that important constraints are satisfied by the new excluded-volume-SRC model. Finally, in Sec.~\ref{summ} we finish our study by exhibiting a brief summary of our main results, and some concluding remarks.

\section{Density dependent vdW model applied to nuclear matter} 
\label{ddvdw}

The idea of converting the classical vdW model into a quantum version with relativistic effects included was originally presented in~\cite{vovchenko2015a,vovchenko2015b,vovchenko2017a,vovchenko2017b}. In principle, such a modification is enough to make the model able to describe the basic phenomenology of the nucleon-nucleon interaction, namely, attractive and repulsive parts, simulated in this approach by the correction in the ideal gas pressure, and the excluded volume, respectively. In the zero temperature regime, the authors have successfully determined numerical values for the two free constants presented in the model ($a$ and $b$) by imposing $B_0=-16$~MeV (binding energy), and $\rho_0=0.16$ fm$^{-3}$ (saturation density) for the symmetric nuclear matter (SNM) case, in which the proton fraction is $y_p=0.5$. This procedure lead to $a \sim 329$ MeV fm$^{3}$ and $b\sim 3.42$ fm${^3}$. Even with the saturation point ensured, the simplest version of the vdW model applied to nuclear matter still contains some issues, such as the value obtained for incompressibility at the saturation density, $K_0\sim 760$~MeV. The modification of the excluded volume mechanism implemented, namely, from the traditional one to the Carnahan-Starling~(CS) procedure~\citep{cs}, decreases this number to $K_0\sim 330$~MeV. However, this value is still not inside the empirical range of $K_0=(240 \pm 20)$~MeV~\citep{shlomo2006,garg2018}, or the one given in~\cite{stone2014}: $K_0=(250 - 315)$~MeV. Despite that, this model and its variations, such as the Clausius-CS model, are capable of reproducing lattice data at finite temperature regime~\citep{vovchenko2017a}. 

In~\cite{lourenco2019} a generalization of the vdW model was proposed. More specifically, possible density dependence in the attractive contribution was taken into account. For the term containing the excluded volume, the aforementioned CS procedure was used as well. In summary, the EoS for energy density and pressure are given, respectively, by
\begin{align}
\epsilon (\rho,y_p) &= \left[1-\rho\mathcal{B}(\rho )\right]
\left(
\epsilon_{\mbox{\tiny kin}}^{\star p} + 
\epsilon_{\mbox{\tiny kin}}^{\star n}
\right)
-\rho^2\mathcal{A}(\rho) 
\nonumber\\
&+ d (2y_p - 1)^2\rho^2,
\label{de-ddvdw}
\end{align}
and
\begin{align}
p(\rho,y_p) &= p^{\star p}_{\mbox{\tiny kin}} + p^{\star n}_{\mbox{\tiny kin}}
- \rho^2\mathcal{A}(\rho)
\nonumber\\
&+ \rho  \Sigma (\rho,y_p) + d(2y_p -1)^2\rho^2,
\label{press-ddvdw}
\end{align}
with $\Sigma (\rho,y_p)=\rho\mathcal{B}'(P^{\star p}_{\mbox{\tiny kin}} + P^{\star n}_{\mbox{\tiny kin}})-\rho^2\mathcal{A}'$ being the rearrangement term with $\mathcal{A}'\equiv d\mathcal{A}/d\rho$, and $\mathcal{B}'\equiv d\mathcal{B}/d\rho$. The kinetic contributions are 
\begin{equation}
\epsilon_{\mbox{\tiny kin}}^{\star p,n}= \frac{\gamma}{2\pi^2}\int^{k_F^{\star p,n}}_0\hspace{-0.5cm}dk\,k^2 \sqrt{k^2+M^{2}},
\label{ekin}
\end{equation}
and
\begin{equation}
p_{\mbox{\tiny kin}}^{\star p,n}= \frac{\gamma}{6\pi^2}
\int^{k_F^{\star p,n}}_0\hspace{-0.5cm}\frac{dk\,k^4}{\sqrt{k^2+M^2}}.
\label{pkin}
\end{equation}
The Fermi momentum of the nucleon of mass $M=939$~MeV and degeneracy factor $\gamma=2$ is related to its respective density as $k_F^{\star p,n}= (6\pi^2 \rho_{ p,n}^\star/\gamma)^{1/3}$, where
\begin{equation}
\begin{split}
\rho^{\star}_p & = \frac{y_p\rho}{1-\rho\mathcal{B}(\rho)} 
= \frac{\rho_p}{1-\rho\mathcal{B}(\rho)}, 
\\
\rho^{\star}_n & =\frac{(1-y_p)\rho}{1-\rho\mathcal{B}(\rho)}
= \frac{\rho_n}{1-\rho\mathcal{B}(\rho)}.
\end{split}
\label{c3_densityod}
\end{equation}
Finally, the density-dependent functions $\mathcal{A}$ and $\mathcal{B}$ are
\begin{equation}
\mathcal{A}(\rho)= \frac{a}{(1+b\rho)^n},
\label{arho}
\end{equation}
and
\begin{equation}
\mathcal{B}(\rho) =\frac{1}{\rho }-\frac{1}{\rho }\exp \left[-\frac{b \rho }{4}\frac{\left(4-\frac{3b\rho}{4}\right)}{\left(1-\frac{b\rho}{4}\right)^2}\right],
\label{brho}
\end{equation}
with this last one determined through the CS approach for the repulsive interaction (excluded volume). It is worth noticing that from this general structure, it is possible to recover the other real gases studied in~\cite{vovchenko2017a} for the $y_p=0.5$ case, as for instance the \mbox{vdW-CS} model itself, by using $n=0$, and the \mbox{Clausius-CS} one, for which $n=1$. The traditional versions of these models regarding the excluded volume method are obtained by making $\mathcal{B}(\rho)\rightarrow b$ in addition. Another formulation involving a vdW model in which induced surface tension is taken into account was implemented in~\cite{sagun,bugaev}.

This new model, named the density-dependent vdW (\mbox{DD-vdW}) model, has shown to preserve causality in a density regime capable of producing mass-radius diagrams consistent with data obtained from the PSR J0348+0432 pulsar~\citep{psrJ0348+0432}, as well as those from the GW170817 neutron-star merger event. It is also compatible with the flow constraint established in~\cite{danielewicz2002}. The four free parameters ($a$, $b$, $d$ and $n$) are determined by imposing specific values for $\rho_0$, $B_0$, $K_0$ and $J$ (symmetry energy at~$\rho_0$). Furthermore, it also produces some clear correlations in SNM as one can see in~\cite{jpgnosso}.

\section{Short-range correlations: CCS-SRC model} 
\label{src}

\begin{figure} 
\centering
\includegraphics[scale=0.3]{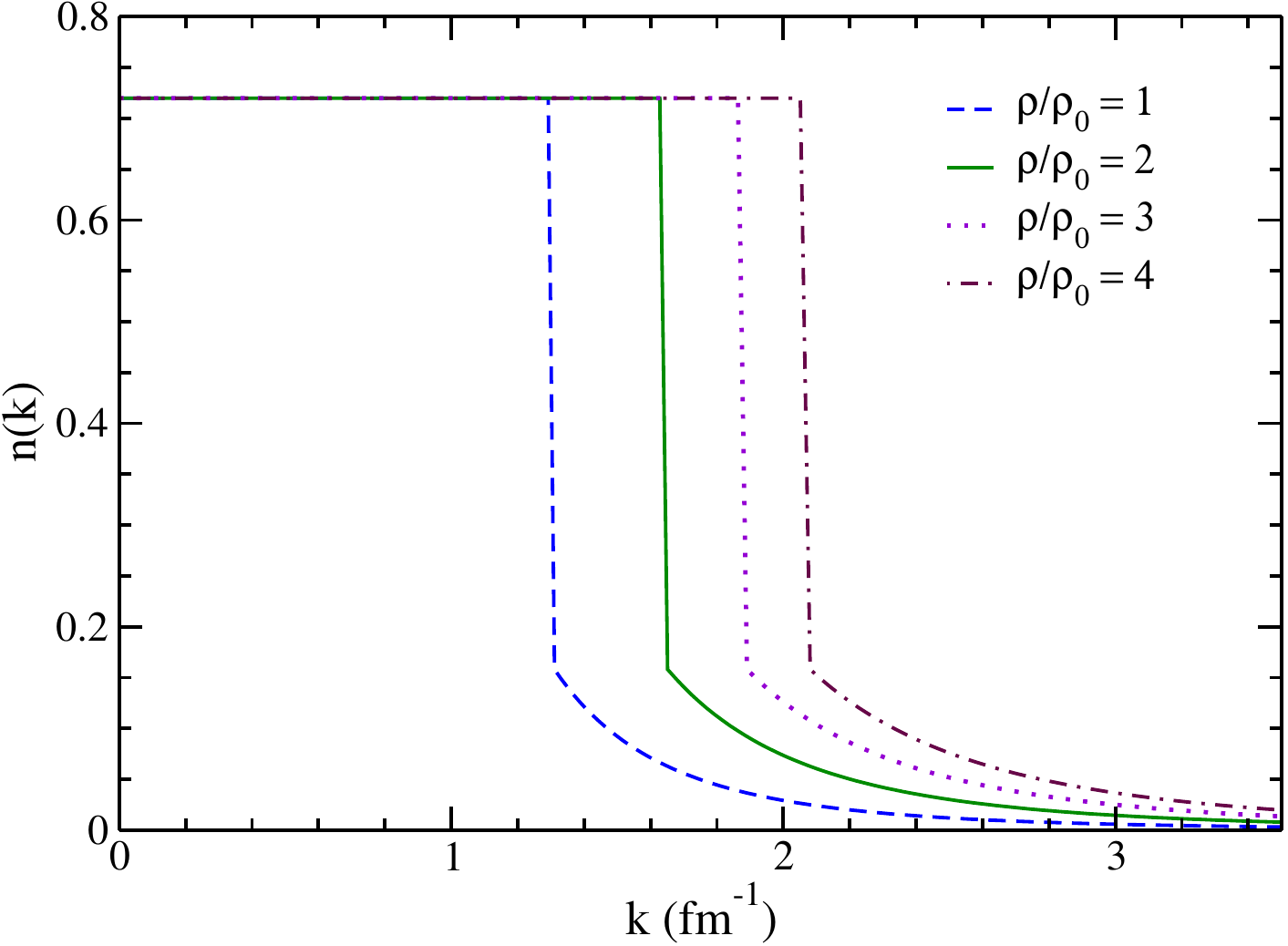}
\caption{Momentum distribution with HMT included for symmetric nuclear matter. Curves for $\rho/\rho_0=1,2,3$ and $4$, with $\rho_0=0.15$ fm$^{-3}$.} 
\label{hmt}
\end{figure}

The inclusion of SRC in hadronic models is performed by modifying the single-nucleon momentum distributions, from the usual Fermi step functions to those encompassing the high-momentum tail (HMT) that read
\begin{eqnarray}
n_{p,n}(k) = \left\{ 
\begin{array}{ll}
\Delta_{p,n}, & 0<k<k_F^{p,n}
\\ \\
C_{p,n}\dfrac{(k_F^{p,n})^4}{k^4}, & k_F^{p,n}<k<\phi_{p,n} k_F^{p,n},
\end{array} 
\right.
\label{htm}
\end{eqnarray}
with $\Delta_{p,n}=1 - 3C_{p,n}(1-1/\phi_{p,n})$, $C_p=C_0[1 - C_1(1-2y_p)]$, $C_n=C_0[1 + C_1(1-2y_p)]$, $\phi_p=\phi_0[1 - \phi_1(1-2y_p)]$ and $\phi_n=\phi_0[1 + \phi_1(1-2y_p)]$. The values $C_0=0.161$, $C_1=-0.25$, $\phi_0 = 2.38$, and $\phi_1=-0.56$ are determined~\citep{baoli2015,baoli2016,cai2016} by taking experimental data concerning $d(e,e',p)$ and two-nucleon knockout reactions, medium-energy photonuclear absorption, as well as by using the normalization condition
\begin{equation}
\frac{1}{\pi^2} \int_0^{\infty}dk\,k^2\,n_{p,n}(k) = \rho_{p,n} = \frac{(k_F^{p,n})^3}{3\pi^2}.
\label{norm}
\end{equation}
Furthermore, the fraction of nucleons in the HMT given by $x^{\mbox{\tiny HMT}}=3C_{p,n}(1-\phi_{p,n}^{-1})$ is also used in this determination, namely, $x^{\mbox{\tiny HMT}}_{\mbox{\tiny SNM}}=28\%$ and $x^{\mbox{\tiny HMT}}_{\mbox{\tiny PNM}}=1.5\%$: numbers obtained for symmetric nuclear matter and pure neutron matter, respectively~\citep{baoli2015,baoli2016,cai2016}. In Fig.~\ref{hmt} we depict the $n(k)$ distribution in SNM for some values of $\rho/\rho_0$. 

Recently, some papers have explored possible modifications in Eq.~(\ref{htm}) and their consequences. In~\cite{baoli-dim}, for instance, it was studied the effect of generalizing $n_{p,n}(k)$ to arbitrary dimensions. In~\cite{baoli-k} on the other hand, the authors investigated three different shapes for the SRC HMT, namely, proportional to $k^4$, $k^6$, and $k^9$. The analysis performed in this paper is based on hard photons emissions due to the reactions \mbox{$^{14}$N+$^{12}$C} and \mbox{$^{48}$Ca+$^{124}$Se} at beam energies around the Fermi energy. From the reactions, they analyzed the yields, angular distribution, and energy spectra of the hard photons, leading them to important conclusions. The first is related to the yields, which increase equally for all different powers of $k$. The second is that the shape of the HMT does not affect the angular distribution of the produced hard photons. In this way, if one looks only at the yields or at the angular distribution, the shape seems not to be relevant. The two first conclusions make the third one the most meaningful. The authors have calculated the effects of the HMT shape in the hard photons spectra, finding that this effect is considerable and should not be ignored. The effects are greater as greater are the energy of the photons. 

Here we use the expression given in Eq.~(\ref{htm}) adapted to the case in which excluded volume effects are implemented in the system, namely, taking $k_F^{p,n} \rightarrow k_F^{\star p,n}$, in order to generate new EoS for the vdW-type model presented before. This procedure leads to generalized thermodynamical quantities, such as energy density and pressure, given respectively by,
\begin{align}
\epsilon (\rho,y_p) &= \left[1-\rho\mathcal{B}(\rho )\right]
\left[
\epsilon_{\mbox{\tiny kin(SRC)}}^{\star p} + 
\epsilon_{\mbox{\tiny kin(SRC)}}^{\star n}
\right]
-\rho^2\mathcal{A}(\rho) 
\nonumber\\
&+ d (2y_p - 1)^2\rho^2,
\label{de-ccssrc}
\end{align}
and
\begin{align}
p(\rho,y_p) &= p^{\star p}_{\mbox{\tiny kin(SRC)}} + p^{\star n}_{\mbox{\tiny kin(SRC)}}
- \rho^2\mathcal{A}(\rho)
\nonumber\\
&+ \rho  \Sigma_{\mbox{\tiny SRC}} (\rho,y_p) + d(2y_p -1)^2\rho^2,
\label{press-ccssrc}
\end{align}
where
\begin{align}
\Sigma_{\mbox{\tiny SRC}}(\rho,y_p)=\rho\mathcal{B}'\left[p^{\star p}_{\mbox{\tiny kin(SRC)}} + p^{\star n}_{\mbox{\tiny kin(SRC)}}\right]-\rho^2\mathcal{A}',
\end{align}
and with modified kinetic terms written as
\begin{align}
\epsilon_{\mbox{\tiny kin(SRC)}}^{\star p,n} &= \frac{\gamma\Delta_{ p,n}}{2\pi^2}\int^{k_F^{\star p,n}}_0\hspace{-0.5cm}dk\,k^2 \sqrt{k^2+M^{2}} 
\nonumber\\
& + \frac{\gamma C_{p,n}(k_F^{\star p,n})^4}{2\pi^2}\int^{\phi_{p,n}k_F^{\star p,n}}_{k_F^{\star p,n}}\hspace{-0.5cm} dk\,\frac{\sqrt{k^2+M^2}}{k^2},
\label{ekinsrc}
\end{align}
and
\begin{align}
p_{\mbox{\tiny kin(SRC)}}^{\star p,n} &= \frac{\gamma\Delta_{p,n}}{6\pi^2}
\int^{k_F^{\star p,n}}_0\hspace{-0.5cm}\frac{dk\,k^4}{\sqrt{k^2+M^2}} 
\nonumber\\
&+ \frac{\gamma C_{p,n}(k_F^{\star p,n})^{4}}{6\pi^2}\int^{\phi_{p,n} k_F^{\star p,n}}_{k_F^{\star p,n}}\hspace{-0.6cm}\frac{dk}{\sqrt{k^2+M^2}},
\label{pkinsrc}
\end{align}
with the normalization condition, now taken as $\int_0^\infty n_{p,n}(k)\,k^2\,dk=\rho^\star_{p,n}=(k_F^{\star p,n})^3/3$, giving the same numbers for $C_0$, $C_1$, $\phi_0$, and $\phi_1$. Furthermore, we consider the CS excluded volume mechanism for the function $\mathcal{B}(\rho)$, Eq.~(\ref{brho}). In the case of the attractive density-dependent function $\mathcal{A}(\rho)$, we make $n=1$ and $b\rightarrow c$ in Eq.~(\ref{arho}), namely,
\begin{equation}
\mathcal{A}(\rho)= \frac{a}{1+c\rho}.
\label{arho-ccs}
\end{equation}
By doing so, we actually assume the three parameters Clausius-CS model applied to the nuclear matter as shown in~\cite{vovchenko2017a,vovchenko-clausius}, but also generalized to include SRC effects. Hereafter we name it as $\mbox{CCS-SRC}$ model. The four free parameters of the model, $a$, $b$, $c$ and $d$, are determined by imposing $\rho_0=0.15$~fm$^{-3}$, $B_0=-16$~MeV, $J=E_{\mbox{\tiny sym}}(\rho_0)=32$~MeV, and some values for $K_0=K(\rho_0,y_p=\frac{1}{2})$. The expression for the incompressibility in SNM is given by
\begin{align}
K(\rho) &= 9\frac{\partial P}{\partial\rho}\Big|_{y_p=\frac{1}{2}} 
= 9[\Sigma_{\mbox{\tiny SRC}}(\rho)+ \rho \Sigma'_{\mbox{\tiny SRC}}(\rho)]
\nonumber\\
&+ \frac{1+\mathcal{B}'\rho^2}{[1-\mathcal{B}(\rho)\rho]^2} K^{\star}_{\mbox{\tiny kin(SRC)}} - 9\rho[2\mathcal{A}(\rho) + \mathcal{A}'\rho],
\label{incomp}
\end{align}
with $\Sigma_{\mbox{\tiny SRC}}(\rho)=\Sigma_{\mbox{\tiny SRC}}(\rho,y_p=1/2)$, 
\begin{align}
\Sigma'_{\mbox{\tiny SRC}}(\rho)&= (\mathcal{B}''\rho + \mathcal{B}')p_{\mbox{\tiny kin(SRC)}}^{\star} +\frac{(1+\mathcal{B}'\rho^2)\mathcal{B}'\rho}{9[1-\mathcal{B}(\rho)\rho]^2} K^{\star}_{\mbox{\tiny kin(SRC)}} 
\nonumber\\
&-\mathcal{A}''\rho^2-2\mathcal{A}'\rho,
\label{sigmaprime}
\end{align}
\begin{align}
K^{\star}_{\mbox{\tiny kin(SRC)}} 
&= \frac{3 \Delta k_F^{\star 2}}{\sqrt{k_F^{\star 2}+ M^2}}
\nonumber\\
&+ 3 C_0 k_F^{\star 2} \Bigg[\frac{\phi_0}{\sqrt{\phi_0^2k_F^{\star 2}+ M^2}}
-\frac{1}{\sqrt{k_F^{\star 2}+ M^2}} 
\nonumber\\
&+ \frac{4}{k_F^{\star}} \ \text{ln} \left(\frac{\phi_0k_F^{\star}+\sqrt{\phi_0^2k_F^{\star 2}+ M^2}}{k_F^{\star} + \sqrt{k_F^{\star 2}+ M^2}} \right) \Bigg],
\end{align}
and $\Delta=1-3C_0(1 - 1/\phi_0)$. For $P_{\mbox{\tiny kin(SRC)}}^{\star}$ shown in Eq.~\eqref{sigmaprime},  we use the expression given in Eq.~\eqref{pkinsrc} with $k_F^{\star p,n}$ replaced by $k_F^{\star}$ and $\gamma=4$.

The symmetry energy reads
\begin{align}
E_{\mbox{\tiny sym}}(\rho) &= \frac{1}{8}\frac{\partial^{2}(\epsilon/\rho)}{\partial y_p^{2}}\Big|_{y_p=\frac{1}{2}} 
= \frac{k_F^{\star 2} }{6 E_F^{\star}}\left[1-3 C_0 \left(1-\frac{1}{\phi_0}\right)\right] 
\nonumber\\
&- 3 C_0 E_F^{\star} \left[C_1 \left(1-\frac{1}{\phi_0}\right) +\frac{\phi_1}{\phi_0}\right]
 \nonumber\\
&-\frac{9 M^4}{8 k_F^{\star 3}} \frac{C_0 \phi_1 (C_1-\phi_1) }{\phi_0}
\Bigg[\frac{2 k_F^{\star}}{M} \left(1+\frac{k_F^{\star 2}}{M^2}\right)^{3/2}
 \nonumber\\
&-\frac{k_F^{\star}}{M} \sqrt{1+\frac{k_F^{\star 2}}{M^2}} 
-\text{arcsinh} \left(\frac{k_F^{\star}}{M}\right)\Bigg]
\nonumber\\
&+\frac{2 C_0 k_F^{\star}(6 C_1+1) }{3}\Bigg[\sqrt{1+\frac{M^2}{k_F^{\star 2}}}-\sqrt{1+\frac{M^2}{k_F^{\star 2} \phi_0^2}}
 \nonumber\\
&+\text{arcsinh}\left(\frac{k_F^{\star} \phi_0}{M}\right)-\text{arcsinh}\left(\frac{k_F^{\star}}{M}\right)\Bigg]
\nonumber\\
&+\frac{3 C_0 k_F^{\star}}{2}  \Bigg[\frac{4 E_F^{\star}}{9 k_F^{\star}}-\frac{k_F^{\star}}{9 E_F^{\star}}+\frac{1}{9} (3 \phi_1+1)^2 \left(\frac{k_F^{\star} \phi_0}{F_F^{\star}}-\frac{2 F_F^{\star}}{k_F^{\star} \phi_0}\right)
 \nonumber\\
&+\frac{2 F_F^{\star} (3 \phi_1-1)}{9 k_F^{\star} \phi_0}\Bigg]
\nonumber\\
&+\frac{C_0 (3 C_1+4)}{3} \left[\frac{F_F^{\star} (3 \phi_1+1)}{\phi_0}-E_F^{\star}\right] + d\rho,
\label{esym}
\end{align}
with $E_F^{\star} = \sqrt{k_F^{\star 2} + M^2}$ and $F_F^{\star} = \sqrt{\phi_0^2 k_F^{\star 2} + M^2}$. 

It is worth mentioning that Eq.~\eqref{incomp} reduces to that one related to the \mbox{DD-vdW} model~\citep{lourenco2019} when SRC are turned off, by taking $\phi_0=1$ and $\phi_1=0$, and when Eq.~\eqref{arho} is used instead of Eq.~\eqref{arho-ccs}. With regard to the symmetry energy, notice that its kinetic part, given by $E_{\mbox{\tiny sym}}^{\mbox{\tiny kin}}(\rho)=E_{\mbox{\tiny sym}}(\rho)-d\rho$, is exactly the same presented in~\cite{baoli2016} for the case in which no excluded volume effects are considered in the system, i.e., for $k_F^{\star}=k_F$. Furthermore, we find $E_{\mbox{\tiny sym}}^{\mbox{\tiny kin}}(\rho_0)=-14.7$ MeV for the kinetic part of the symmetry energy at the saturation density. This value is compatible with respective numbers obtained in~\cite{baoli2016} from a nonlinear relativistic mean-field (RMF) model, and from a nonrelativistic calculation, both including SRC effects. 

\begin{figure} 
\centering
\includegraphics[scale=0.33]{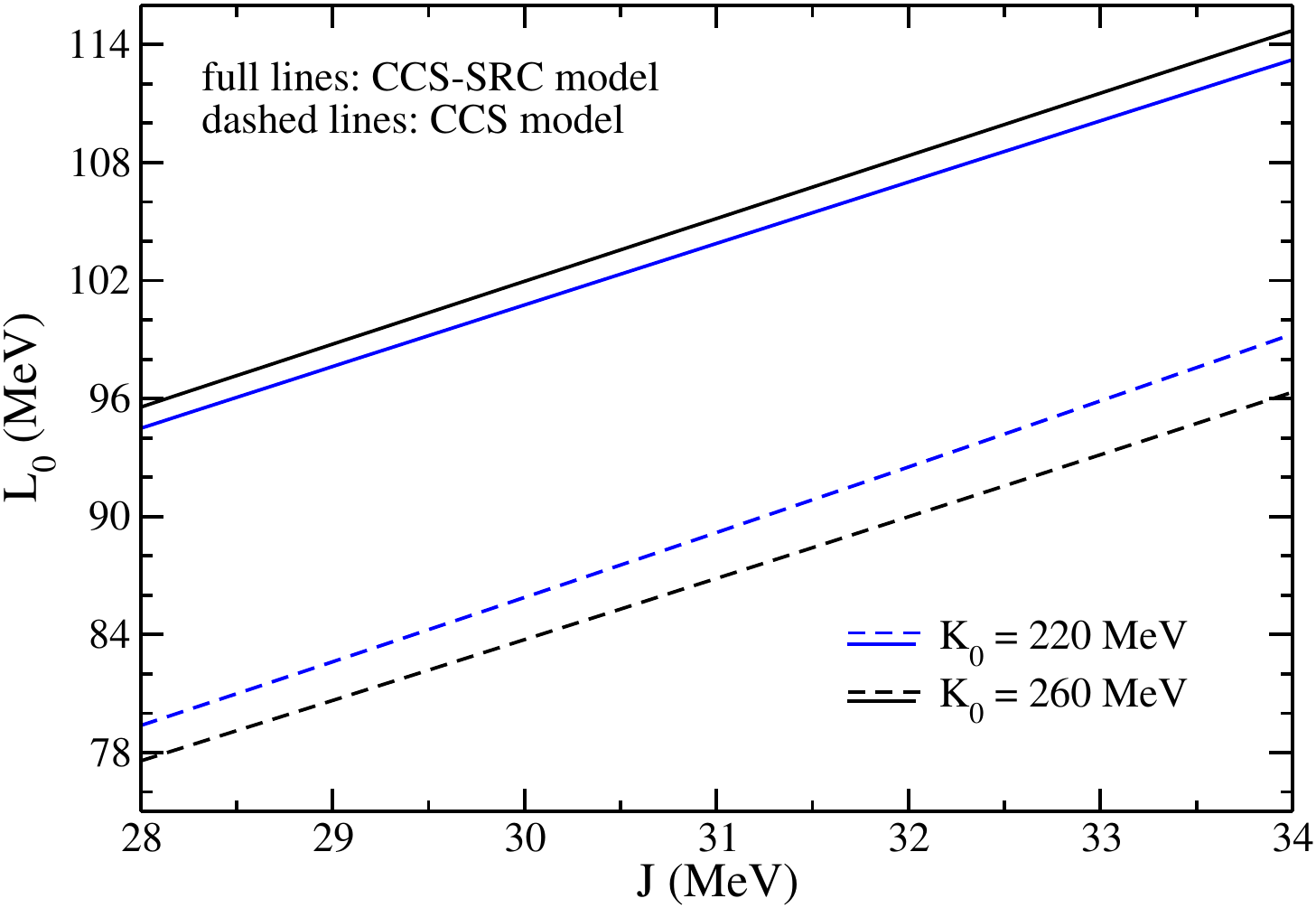}
\caption{$L_0$ as a function of $J$ for the CCS model with (full lines) and without (dashed lines) SRC included. Curves constructed by using $\rho_0=0.15$~fm$^{-3}$, $B_0=-16$~MeV.}  
\label{jl}
\end{figure}

For the sake of completeness, we also investigate how the symmetry energy and its slope, obtained through $L=3\rho(\partial E_{\mbox{\tiny sym}}/\partial\rho)$, correlates with each other (both quantities evaluated at the saturation density: $J$ and $L_0=L(\rho_0)$). Such a relationship is depicted in Fig.~\ref{jl}.
From the figure, it is verified a strong linear correlation between these quantities, in accordance with many other approaches performed in the literature, as can be seen, in~\cite{jl1,jl2,jl3}, for example. Another feature exhibited in the figure is that SRC significantly increases the values of $L_0$ for the same $J$. It is also observed that there is no big impact in $L_0$ for $K_0$ changing in the range of $K_0=(240\pm 20)$~MeV.

\subsection{Applications in SNM and stellar matter} 

\begin{figure} 
\centering
\includegraphics[scale=0.32]{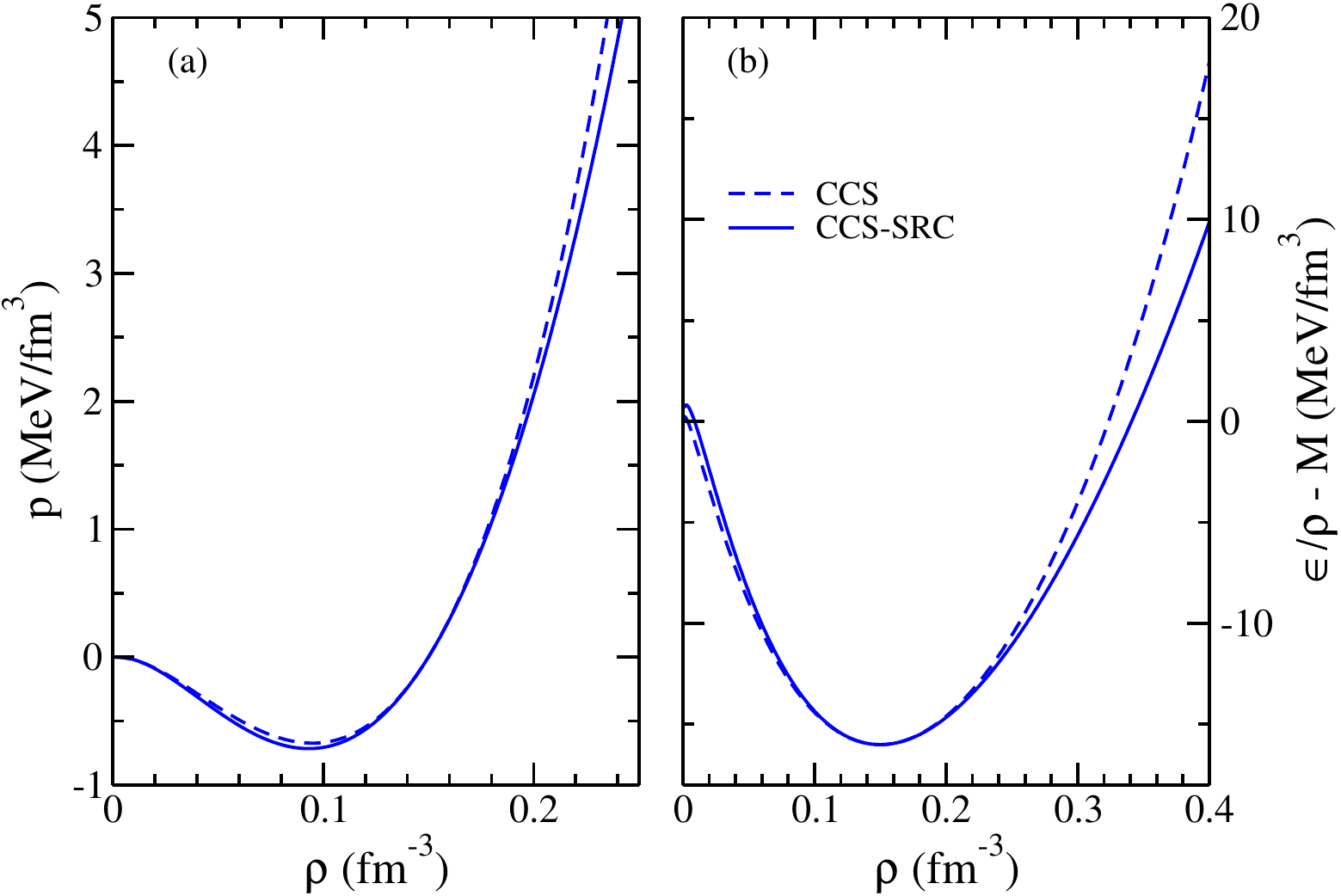}
\caption{CCS model with (full lines) and without (dashed lines) SRC included: (a) pressure and (b) energy per particle as a function of the density. Curves for symmetric nuclear matter with $\rho_0=0.15$~fm$^{-3}$, $B_0=-16$~MeV, and $K_0=240$~MeV.}  
\label{p_exrho}
\end{figure}

\begin{figure} 
\centering
\includegraphics[scale=0.33]{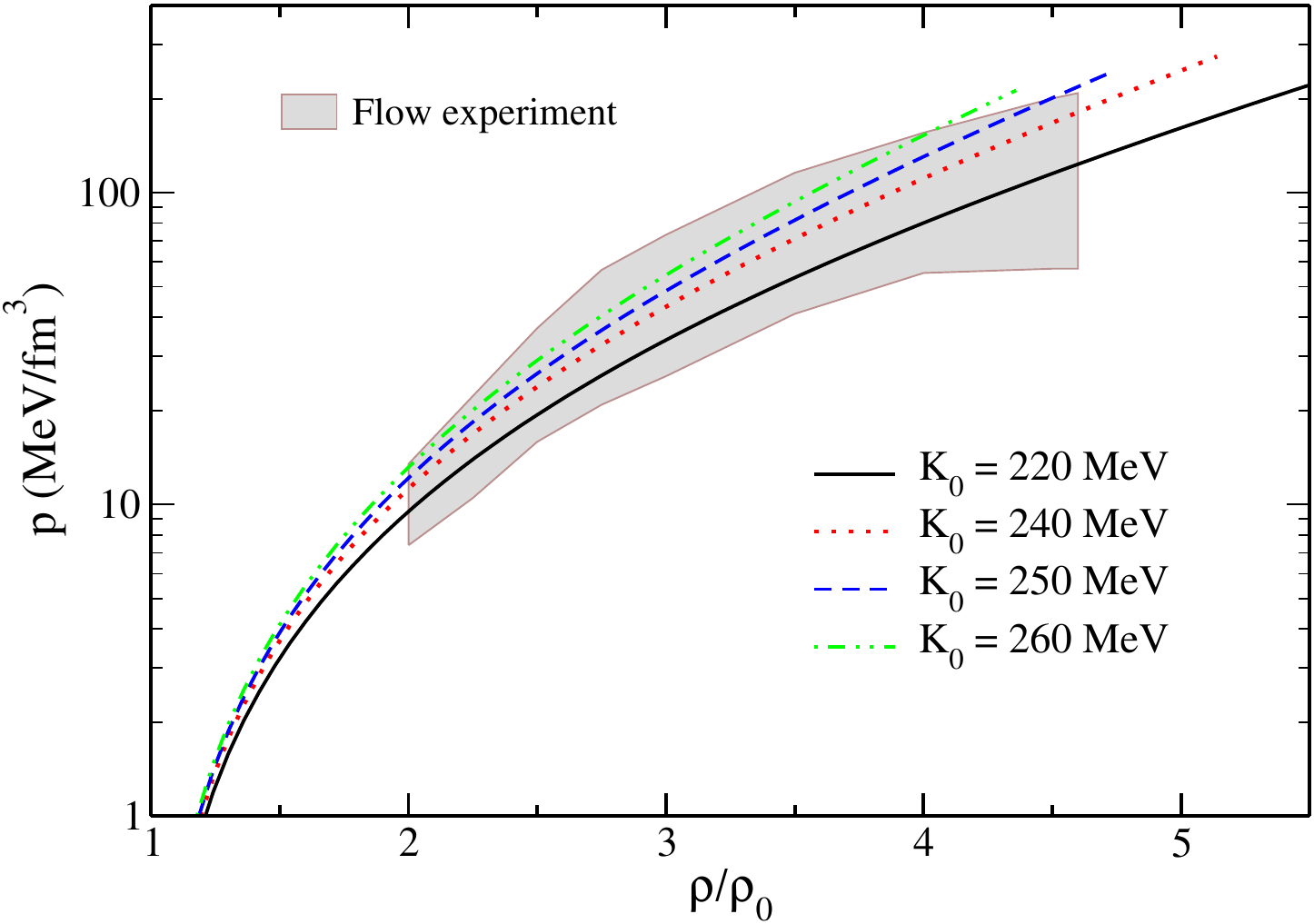}
\caption{Pressure versus $\rho/\rho_0$ for different parametrizations of the $\mbox{CCS-SRC}$ model. Curves for symmetric nuclear matter with $\rho_0=0.15$~fm$^{-3}$ and $B_0=-16$~MeV. Band: flow constraint extracted from~\citep{danielewicz2002}.} 
\label{press-high}
\end{figure}

\begin{figure} 
\centering
\includegraphics[scale=0.33]{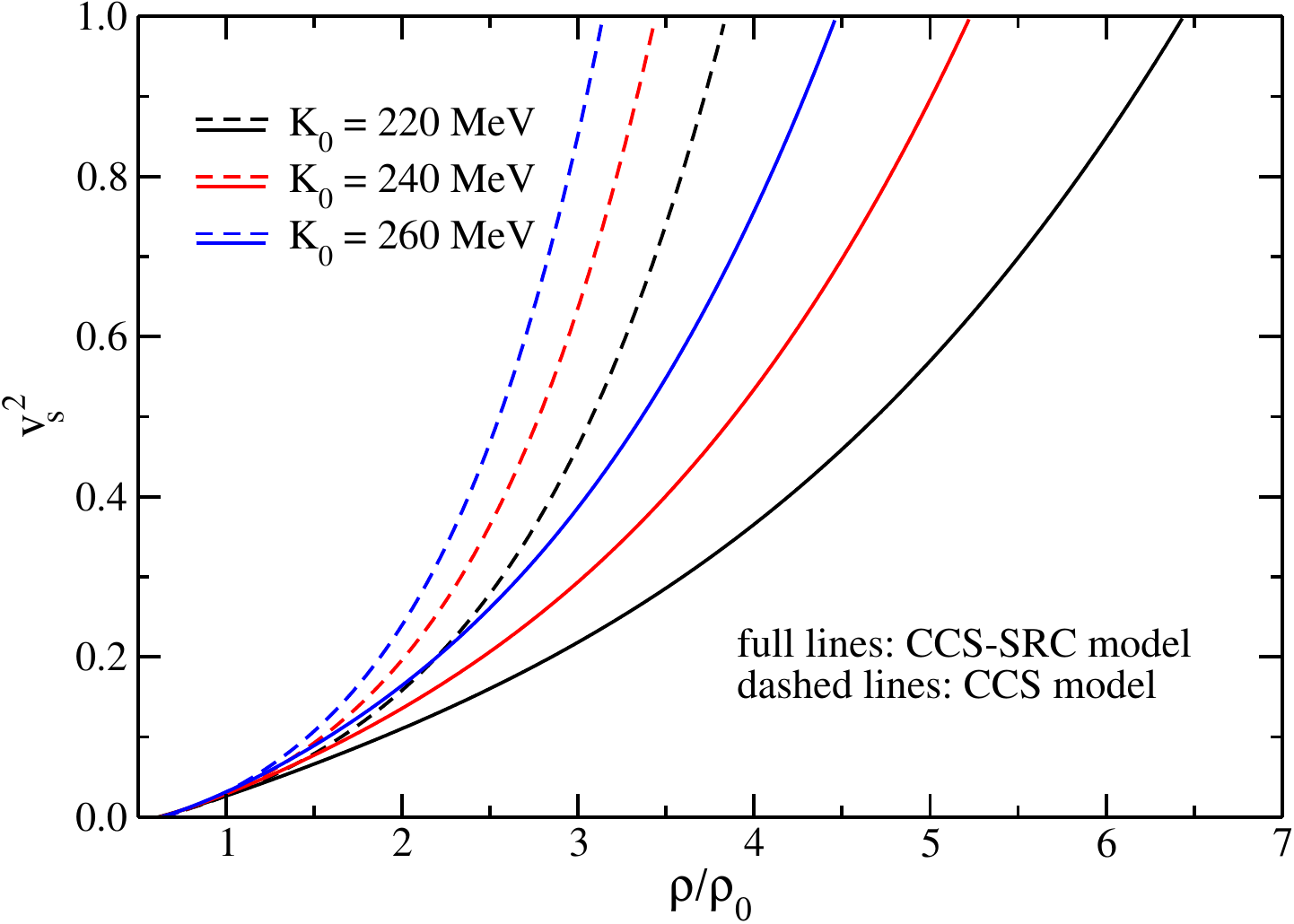}
\caption{Squared sound velocity as a function of $\rho/\rho_0$ for different parametrizations of the CCS model with (full lines) and without (dashed lines) SRC included. Curves for symmetric nuclear matter with $\rho_0=0.15$~fm$^{-3}$ and $B_0=-16$~MeV.}  
\label{vs2}
\end{figure}

We show in Figs.~\ref{p_exrho}{\color{blue}a} and~\ref{p_exrho}{\color{blue}b} the effect of the SRC applied to the $\mbox{CCS-SRC}$ model in the energy per particle and pressure of the system in SNM.
From these figures, we notice that SRC mainly affects such thermodynamical quantities especially for densities greater than $0.2$~fm$^{-3}$, approximately. In this case, it is important to verify the results of the model regarding the high-density regime. For this purpose, we also investigate how it behaves against the so-called flow constraint. It is based on the study performed in~\cite{flow} in which limits on the pressure of SNM (zero temperature case) at high densities were established from experimental data related to the motion of ejected matter in energetic nucleus–nucleus collisions, more specifically, particle flow in the collisions of $^{197}\mbox{Au}$ nucleus at incident kinetic energy per nucleon running from about $0.15$~GeV to $10$~GeV. The comparison of the model with this constraint is displayed in Fig.~\ref{press-high}.
It is verified that parametrizations of the $\mbox{CCS-SRC}$ model constructed by fixing $K_0$ in the range of $K_0=(240\pm 20)$~MeV~\citep{garg2018} are completely in agreement with the band provided by the flow constraint. All these curves were generated in a density range that ensures causality to the system. In the case of excluded volume models, like the one we are presenting here, nucleons are treated as finite-size objects and therefore a suitable Lorentz contraction should be taken into account for relativistic frameworks in order to avoid causality violation for any density~\citep{bugaev08}. An alternative to this procedure is the implementation of the CS excluded volume treatment, since this mechanism effectively produces an excluded volume depending on the density, more specifically, as a decreasing function. In the case of the model proposed in this work, we verify that SRC moves the density in which causality is broken to higher values in comparison with the model without this phenomenology implemented. This feature is observed in Fig.~\ref{vs2}.

We also investigate the capability of the $\mbox{CCS-SRC}$ model in describing stellar matter in general, and some recent astrophysical observations in particular. In order to do that, it is necessary to solve the Tolman-Oppnheimer-Volkoff (TOV) equations~\citep{tov39,tov39a}, given by $dP(r)/dr=-[\varepsilon(r) + P(r)][m(r) + 4\pi r^3P(r)]/[r^2g(r)]$ and $dm(r)/dr=4\pi r^2\varepsilon(r)$, where $g(r)=1-2m(r)/r$. The solution of these equations is constrained to $P(0)=P_c$ (central pressure) and $m(0) = 0$, with the conditions $P(R) = 0$ and $m(R)=M_{\mbox{\tiny NS}}$ satisfied at the star surface. Here $R$ defines the radius of the respective neutron star of mass $M_{\mbox{\tiny NS}}$. We impose the NS core as described by the EoS obtained from the $\mbox{CCS-SRC}$ model. For the outer crust, on the other hand, we use the EoS constructed by Baym, Pethick, and Sutherland (BPS)~\citep{bps} in a density range of $6.3\times10^{-12}\,\mbox{fm}^{-3}\leqslant\rho_{\mbox{\tiny outer}}\leqslant2.5\times10^{-4}\,\mbox{fm}^ {-3}$. Finally, for the inner crust of the NS, we use the polytropic form of $P(\varepsilon)=\alpha+\beta\varepsilon^{4/3}$ from $2.5\times10^{-4}\,\mbox{fm}^ {-3}$ to the transition density, in our case obtained through the thermodynamical method~\citep{tm1,tm2,tm4}. 

The total energy density and total pressure of the system composed of protons, neutrons, electrons, and muons are written as
\begin{eqnarray}
\varepsilon = \epsilon + \frac{\mu_e^4}{4\pi^2} + \frac{1}{\pi^2}\int_0^{\sqrt{\mu_\mu^2(\rho_e)-m^2_\mu}}\hspace{-1cm}dk\,k^2(k^2+m_\mu^2)^{1/2}\quad
\label{det}
\end{eqnarray}
and
\begin{eqnarray} 
P = p + \frac{\mu_e^4}{12\pi^2}
+ \frac{1}{3\pi^2}\int_0^{\sqrt{\mu_\mu^2-m^2_\mu}}\hspace{-0.5cm}\frac{dk\,k^4}{(k^2+m_\mu^2)^{1/2 } },
\label{presst}
\end{eqnarray}
where, by chemical equilibrium and charge neutrality conditions, both imposed in an NS, one has $\mu_n - \mu_p = \mu_e$ and $\rho_p - \rho_e = \rho_\mu$, with $\mu_e=(3\pi^2\rho_e)^{1/3}$, \mbox{$\rho_\mu=[(\mu_\mu^2 - m_\mu^2)^{3/2}]/(3\pi^2)$}, and  $\mu_\mu=\mu_e$, for $m_\mu=105.7$~MeV (muon mass) and massless electrons. $\epsilon$, and $p$ are determined from the $\mbox{CCS-SRC}$ model, as well as the chemical potentials for, namely,
\begin{align}
\mu_{p,n} &= \frac{\partial\epsilon}{\partial\rho_{p,n}}
= \Delta_{p,n} \mu^{\star p,n}_{\mbox{\tiny kin}} + \mu^{\star p,n}_{\mbox{\tiny kin(SRC)}} 
\nonumber\\
&+ \mathcal{B}(\rho)[P^{\star p}_{\mbox{\tiny kin(SRC)}} + P^{\star n}_{\mbox{\tiny kin(SRC)}}]
\nonumber\\
&+\Sigma_{\mbox{\tiny SRC}}(\rho,y_p) -2\mathcal{A}(\rho)\rho \pm 2d (2y_p -1) \rho 
\label{mupn}
\end{align}
for protons (upper sign) and neutrons (lower sign), with
\begin{align}
&\mu^{\star p,n}_{\mbox{\tiny kin(SRC)}} = 3C_{p,n}\Bigg[\mu^{\star p,n}_{\mbox{\tiny kin}}-\frac{(\phi_{p,n}^2 k_{F p,n}^{\star 2} + M^{2})^{1/2}}{\phi_{p,n}} \Bigg]
\nonumber\\
&+4C_{p,n}k_{F p,n}^{\star} \text{ln}\Bigg[\frac{\phi_{p,n}^2 k_{F p,n}^{\star 2} + (\phi_{p,n}^2 k_{F p,n}^{\star 2} + M^{2})^{1/2}}{ k_{F p,n}^{\star}+( k_{F p,n}^{\star 2} +M^{2})^{1/2}} \Bigg],
\end{align}
and $\mu^{\star p,n}_{\mbox{\tiny kin}}=(k_{F p,n}^{\star 2} + M^2)^{1/2}$. Notice that Eqs.~\eqref{mupn} reduce to the chemical potentials of the \mbox{DD-vdW} model when SRC are turned off ($\phi_0=1$ and $\phi_1=0$ case). Furthermore, in the case of no excluded volume implemented in the model, i.e., for $\mathcal{B}(\rho)\rightarrow 0$, the first two terms of Eqs.~\eqref{mupn} become identical to ones related to the relativistic model studied in~\cite{lucas}, for $M\rightarrow M^*$, see Eq.~6 to~8 of that reference.

\begin{figure} 
\centering
\includegraphics[scale=0.33]{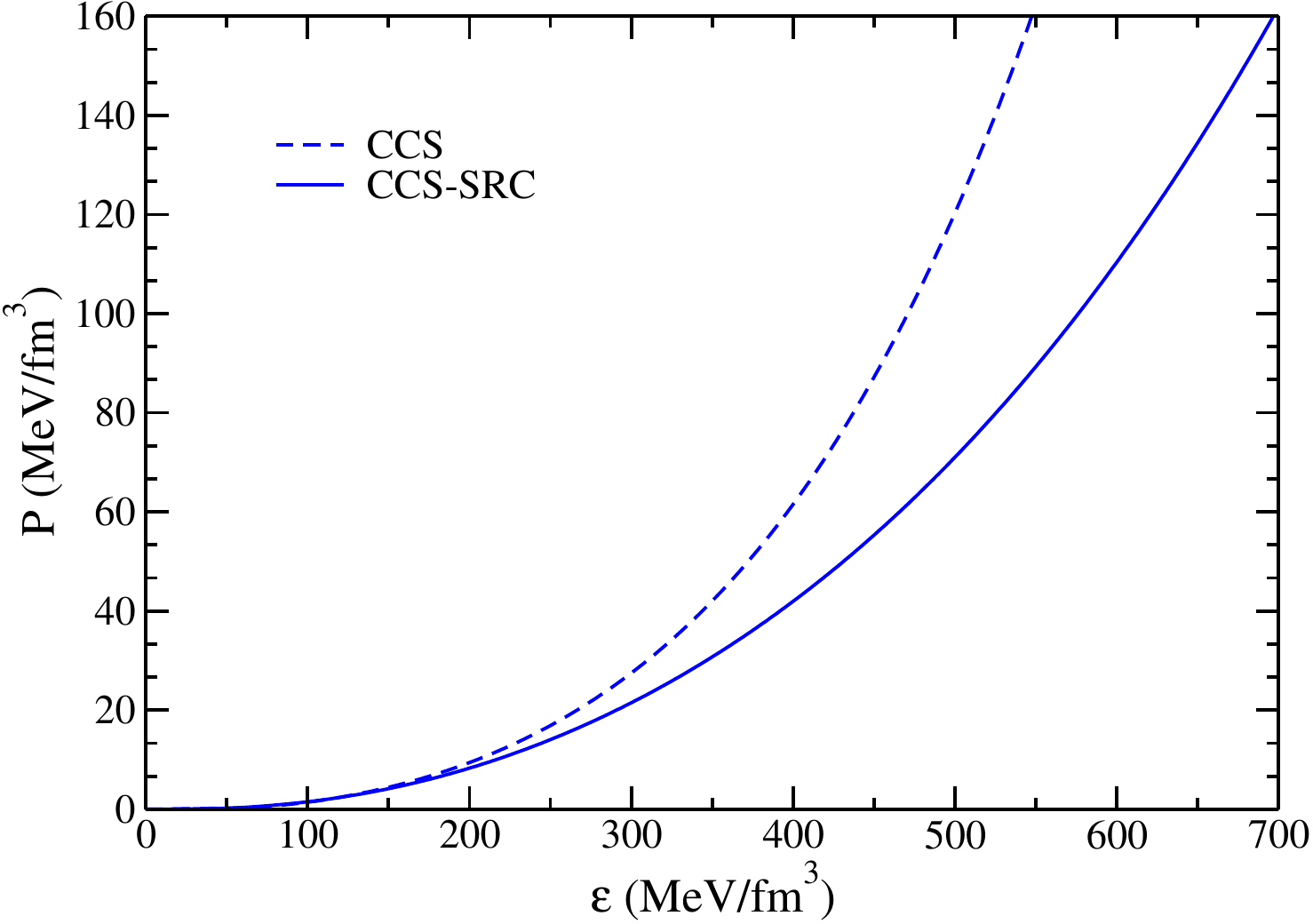}
\caption{Total pressure vs total energy for the CCS model with (full lines) and without (dashed lines) SRC included. Curves for stellar matter with $\rho_0=0.15$~fm$^{-3}$, $B_0=-16$~MeV, $K_0=240$~MeV, and $J=32$~MeV.}  
\label{peccs}
\end{figure}

\begin{figure} 
\centering
\includegraphics[scale=0.325]{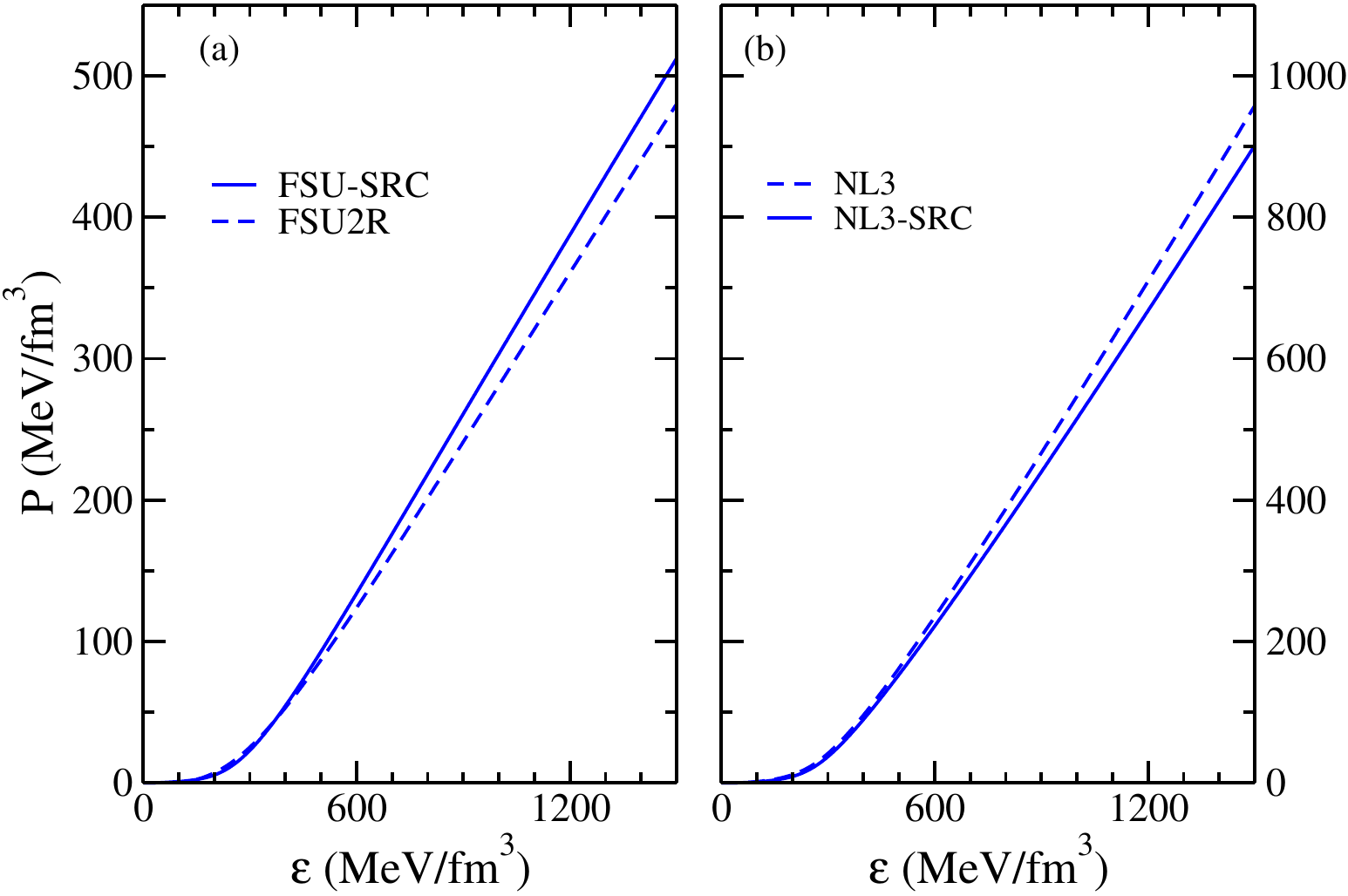}
\caption{Total pressure as a function of total energy density (stellar matter) for the (a) FSU and (b) NL3 parametrizations with (full lines) and without (dashed lines) SRC included.} 
\label{permf}
\end{figure}

\begin{figure} 
\centering
\includegraphics[scale=0.08,trim=0 6cm 10cm 10cm,clip=true]{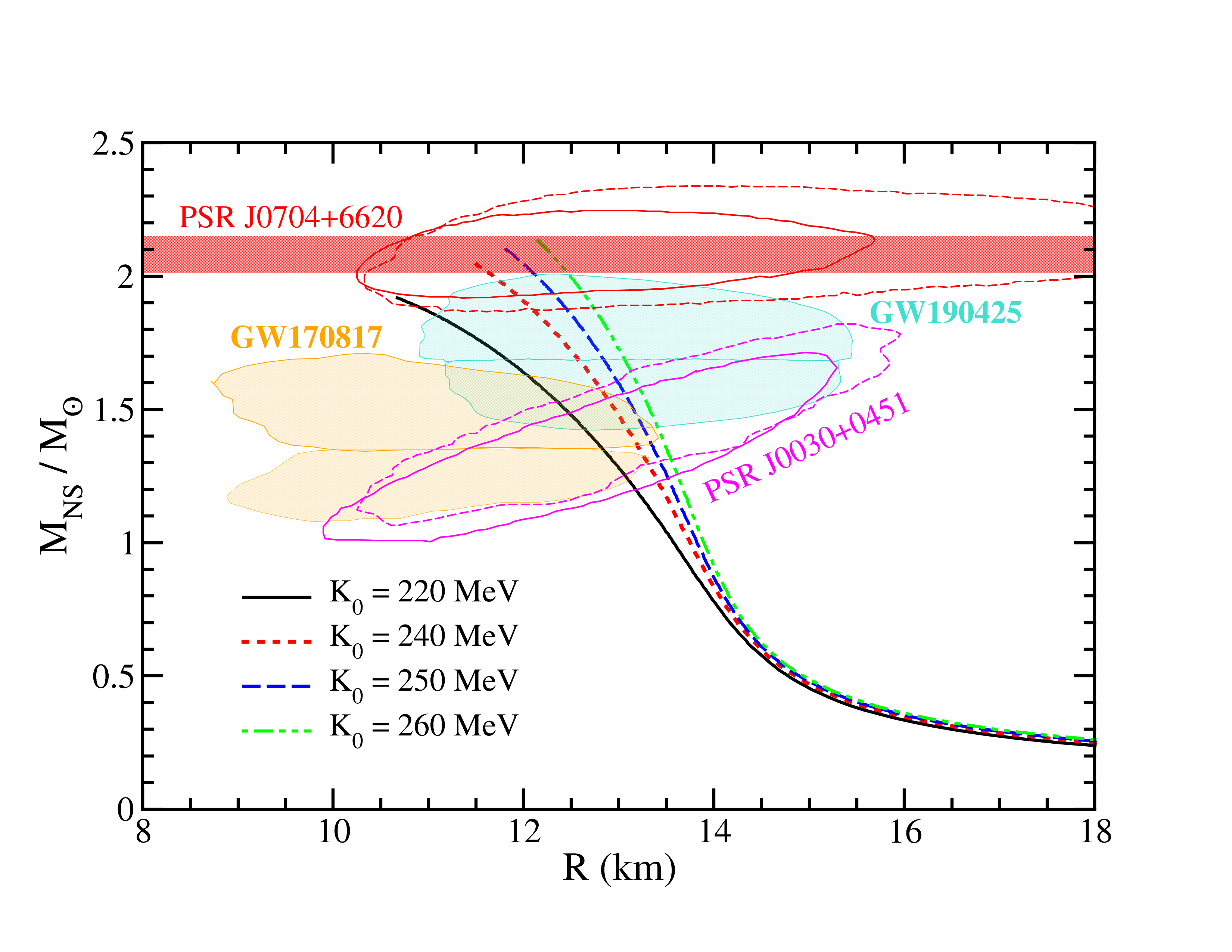}
\caption{Mass-radius diagrams constructed from the \mbox{CCS-SRC} model with different values of $K_0$. The contours are related to data from the NICER mission, namely, PSR~J0030+0451~\citep{Riley_2019,Miller_2019} and PSR~J0740+6620~\citep{Riley_2021,Miller_2021}, the GW170817~\citep{Abbott_2017,Abbott_2018} and the GW190425 events~\citep{Abbott_2020-2}, all of them at $90\%$ credible level. The red horizontal lines are also related to the PSR~J0740+6620 pulsar~\citep{Fonseca_2021}.} 
\label{mrdata}
\end{figure}

Before presenting the outcomes of the model concerning the mass-radius diagrams, we first discuss the effect of SRC in the EoS used as input to the TOV equations, by analyzing the outcomes presented in Fig.~\ref{peccs}.
As already mentioned, SRC move the break of causality to higher densities, or equivalently, to higher energy densities in the case of the data shown in the figure. Moreover, one can also notice that SRC make softer the EoS since the pressure is lower for the same value of $\epsilon$ in comparison with the case in which no SRC are included. This is not the case for RMF models that present quartic interaction in the vector field $\omega_\mu$, i.e., a term given by $C_\omega(\omega_\mu\omega^\mu)^2$ in its Lagrangian density, where $C_\omega$ is a constant free parameter, namely~\citep{rev3,dutra2014},
\begin{align}
&\mathcal{L} = \overline{\psi}(i\gamma^\mu\partial_\mu - M)\psi 
+ g_\sigma\sigma\overline{\psi}\psi 
- g_\omega\overline{\psi}\gamma^\mu\omega_\mu\psi
\nonumber \\ 
&- \frac{g_\rho}{2}\overline{\psi}\gamma^\mu\vec{\rho}_\mu\vec{\tau}\psi
+\frac{1}{2}(\partial^\mu \sigma \partial_\mu \sigma - m^2_\sigma\sigma^2)
- \frac{A}{3}\sigma^3 - \frac{B}{4}\sigma^4 
\nonumber\\
&-\frac{1}{4}F^{\mu\nu}F_{\mu\nu} 
+ \frac{1}{2}m^2_\omega\omega_\mu\omega^\mu 
+ C_\omega(\omega_\mu\omega^\mu)^2 -\frac{1}{4}\vec{B}^{\mu\nu}\vec{B}_{\mu\nu} 
\nonumber \\
&+ \frac{1}{2}\alpha'_3g_\omega^2 g_\rho^2\omega_\mu\omega^\mu
\vec{\rho}_\mu\vec{\rho}^\mu + \frac{1}{2}m^2_\rho\vec{\rho}_\mu\vec{\rho}^\mu.
\label{dlag}
\end{align}
For models with this structure, it is verified that SRC make stiffer the EoS (the pressure is higher for the same energy density). For instance, we display in Fig.~\ref{permf}{\color{blue}a} this finding for the FSU2R parametrization~\citep{fsu2r} with and without SRC.
For the construction of these curves, the bulk parameters were kept the same for both approaches (with and without SRC) as well as the value of the constant $C_\omega$, the procedure also adopted in~\cite{baoli2016,lucas}. Despite this result, it is worth mentioning that SRC can also soften the EoS even for RMF models. This is the case for parametrizations with $C_\omega=0$. As an example, we plot in Fig.~\ref{permf}{\color{blue}b} total pressure as a function of total energy density for the NL3~\citep{nl3,silva} parametrization, for which there is no quartic self-interaction in the repulsive vector channel. As we see, the effect of including SRC is exactly the opposite of that verified for the FSU2R parametrization, but the same as the one presented by the \mbox{CCS-SRC} model. It is known that hadronic models with stiffer EoS produce more massive neutron stars. This is a direct consequence of introducing SRC in RMF models with $C_\omega\ne 0$, as verified in~\cite{baoli2016,lucas,dmnosso1,dmnosso2}, for instance. For the case of models with softer EoS, the opposite is expected. In our case, despite SRC generating softer EoS, we still find possible parametrizations of the \mbox{CCS-SRC}  model capable of reproducing recent astrophysical observational data, as presented in Fig.~\ref{mrdata}.
Notice that the model produces mass-radius diagrams in agreement with the following astrophysical constraints: gravitational waves data related to the GW170817~\citep{Abbott_2017,Abbott_2018} and GW190425~\citep{Abbott_2020-2} events, some of them provided by the LIGO and Virgo Collaboration; data from the NICER mission regarding the pulsars PSR~J0030+0451~\citep{Riley_2019,Miller_2019} and PSR~J0740+6620~\citep{Riley_2021,Miller_2021}; and data from the latter pulsar extracted from~\cite{Fonseca_2021}.

\begin{figure} 
\centering
\includegraphics[scale=0.33]{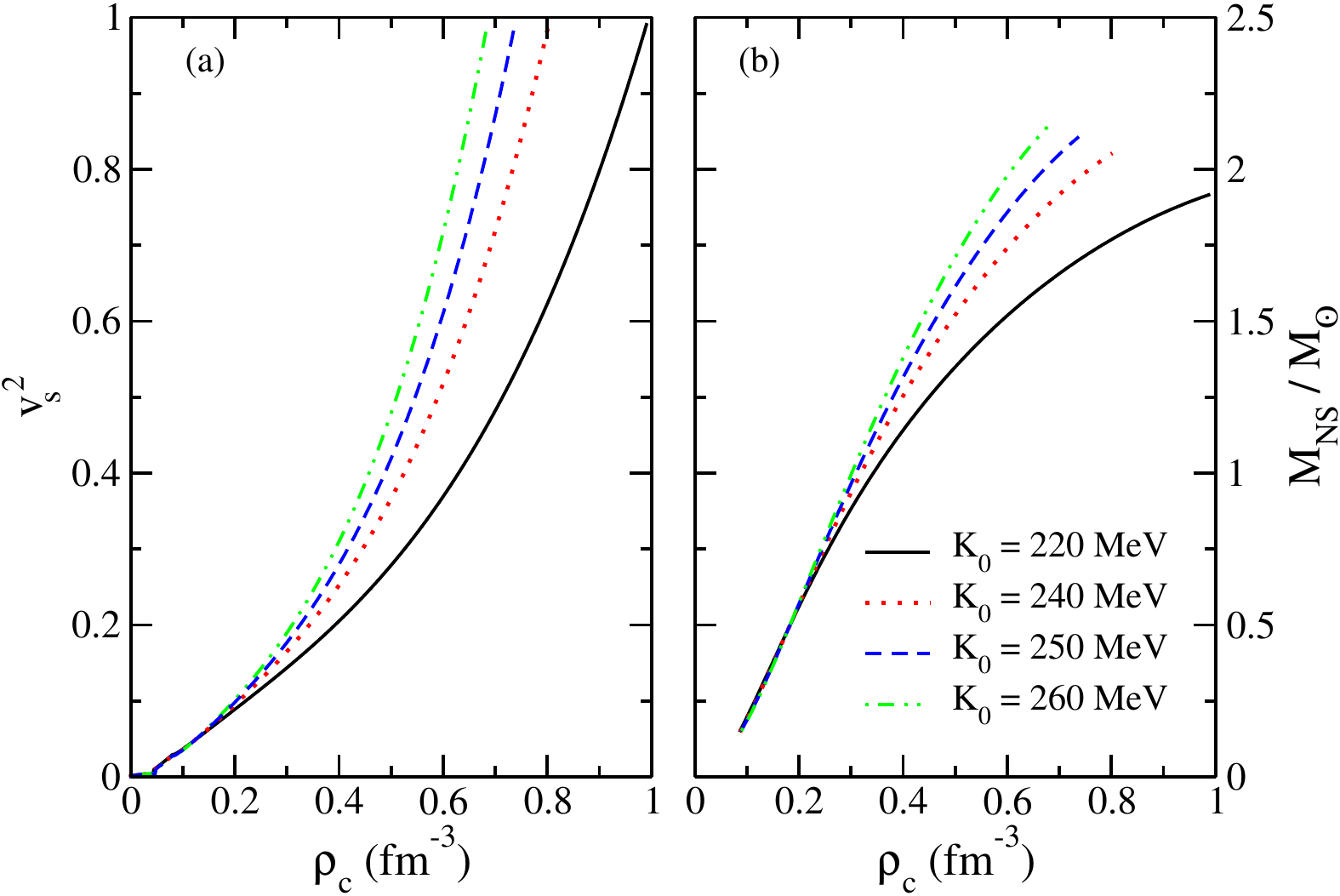}
\caption{(a) Squared sound velocity for beta-equilibrated matter, and (b) stellar mass in units of $M_\odot$, both as a function of the central density for the \mbox{CCS-SRC} model. All curves constructed by using $\rho_0=0.15$~fm$^{-3}$, $B_0=-16$~MeV, $J=32$~MeV, and different values of $K_0$.} 
\label{mr-vs2}
\end{figure}

For the sake of completeness, we present in Fig.~\ref{mr-vs2}{\color{blue}{b}} the plot of the stellar mass as a function of the central density for the different \mbox{CCS-SRC} parametrizations used here. 
In addition, we show in Fig.~\ref{mr-vs2}{\color{blue}{a}} the squared sound velocity for beta-equilibrated matter, $v_s^2=\partial P/\partial\varepsilon$, also as a function of the density. By comparing the results of both panels, it is possible to confirm that a break of causality is not observed for the configurations of the stars generated by the model. 

\begin{figure} 
\centering
\includegraphics[scale=0.33]{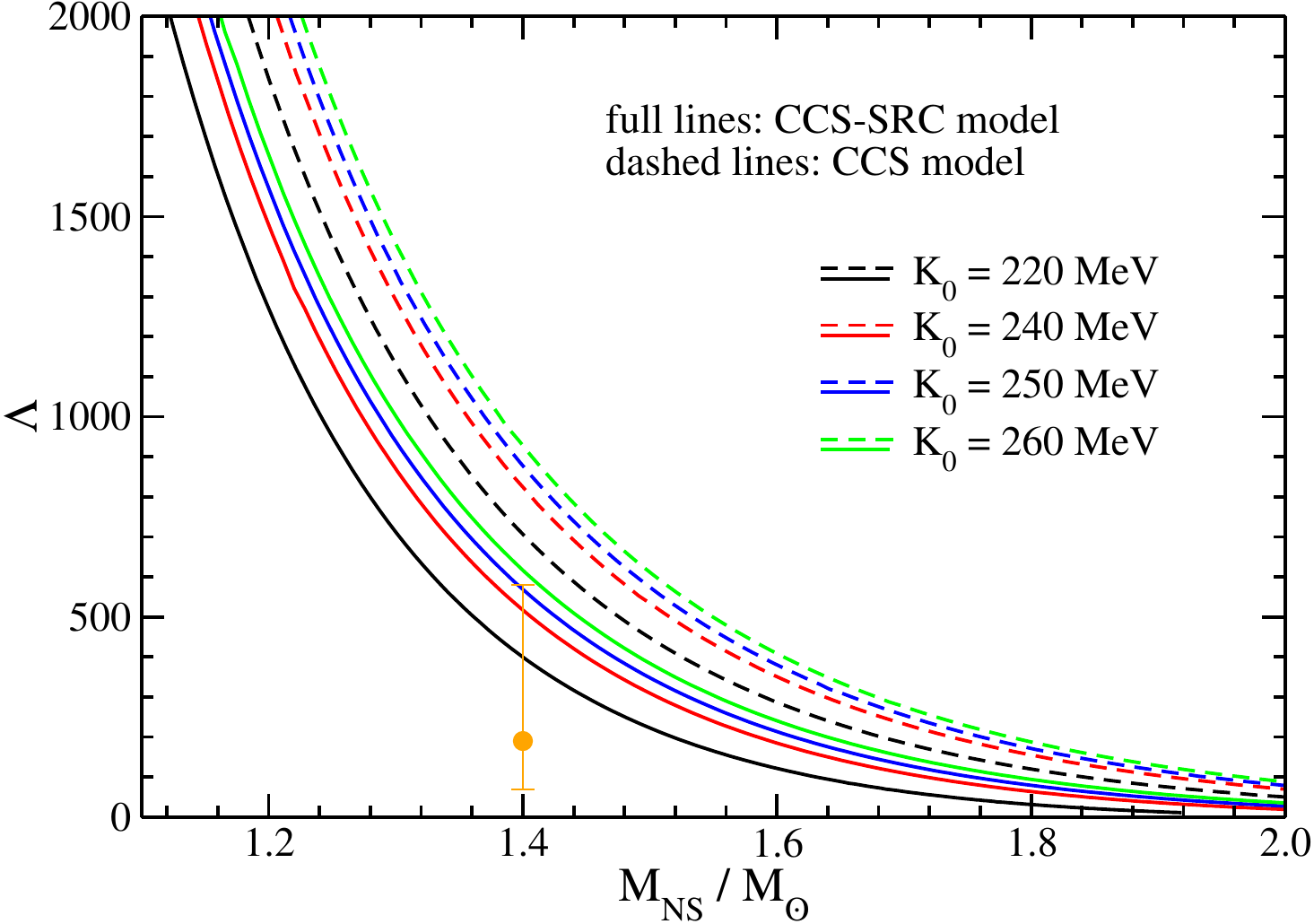}
\caption{$\Lambda$ versus $M_{\mbox{\tiny NS}}/M_\odot$ for the \mbox{CCS} model with $\rho_0=0.15$~fm$^{-3}$, $B_0=-16$~MeV, $J=32$~MeV, and different values of $K_0$ with (full lines) and without (dashed lines) SRC included. Full circle with error bars: result of $\Lambda_{1.4}=190^{+390}_{-120}$ obtained in~\citep{Abbott_2018}.}
\label{lambda}
\end{figure}

\begin{figure} 
\centering
\includegraphics[scale=0.33]{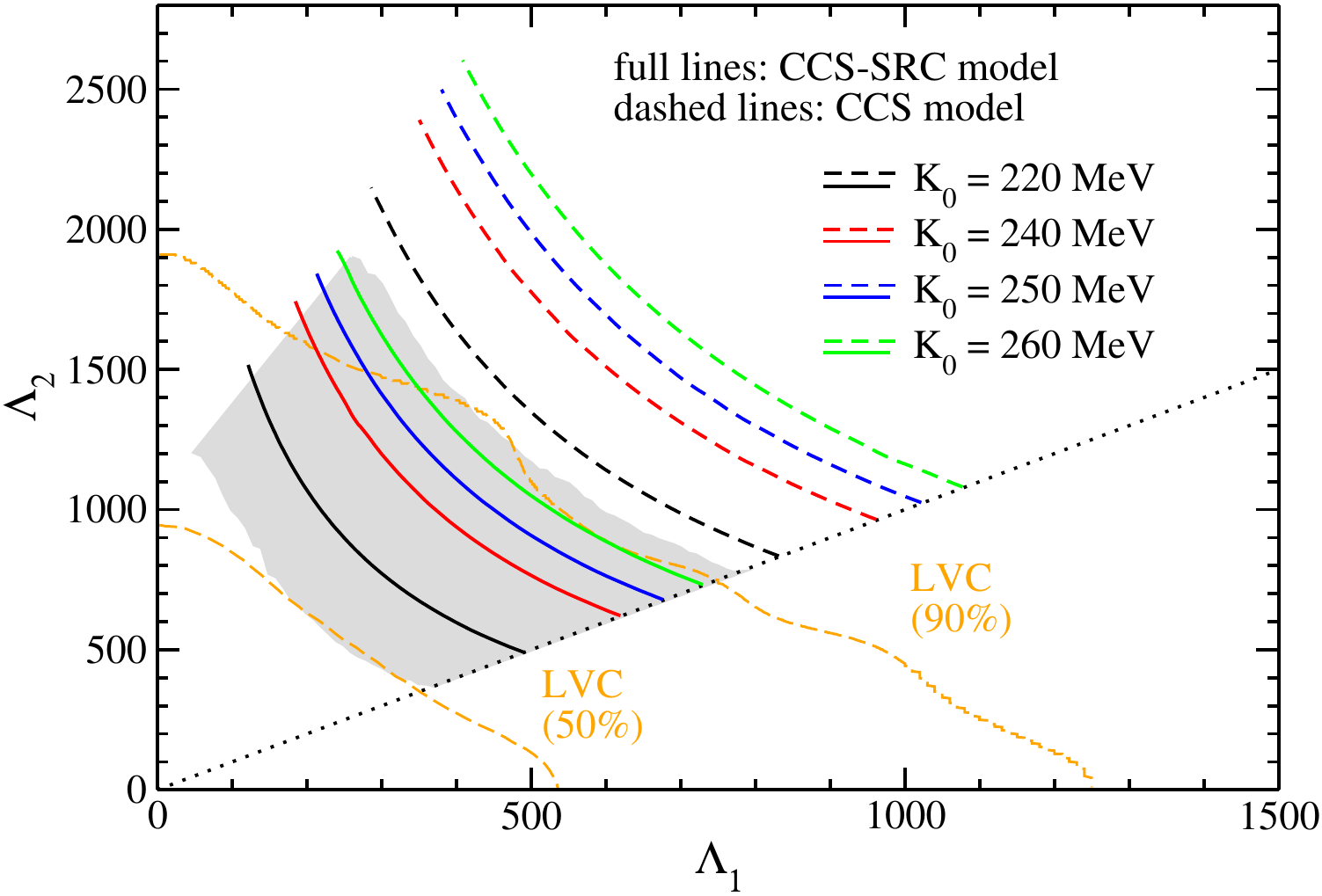}
\caption{$\Lambda_2$ versus $\Lambda_1$ for the \mbox{CCS} model with $\rho_0=0.15$~fm$^{-3}$, $B_0=-16$~MeV, $J=32$~MeV, and different values of $K_0$ with (full lines) and without (dashed lines) SRC included. The orange dashed lines correspond to the $90\%$ and $50\%$ confidence limits given by the LIGO and Virgo Collaboration~(LVC)~\citep{Abbott_2018}. The gray band represents the results obtained through the relativistic mean field models studied in~\citep{defrmf}.}
\label{l1l2}
\end{figure}

We also verify the results obtained through the model with regard to the dimensionless tidal deformability. This quantity is defined as $\Lambda = 
2k_2/(3C^5)$, with $C=M_{\mbox{\tiny NS}}/R$, and the second Love number given by
\begin{eqnarray}
&k_2& = \frac{8C^5}{5}(1-2C)^2[2+2C(y_R-1)-y_R]\nonumber\\
&\times&\Big\{2C [6-3y_R+3C(5y_R-8)] \nonumber\\
&+& 4C^3[13-11y_R+C(3y_R-2) + 2C^2(1+y_R)]\nonumber\\
&+& 3(1-2C)^2[2-y_R+2C(y_R-1)]{\rm ln}(1-2C)\Big\}^{-1},\qquad
\label{k2}
\end{eqnarray}
with $y_R=y(R)$. The quantity $y(r)$ is determined from the solution of the differential equation $r(dy/dr) + y^2 + yF(r) + r^2Q(r)=0$, solved simultaneously with the TOV ones. The expressions for the functions $F(r)$ and $Q(r)$ are
\begin{eqnarray}
F(r) &=& \frac{1 - 4\pi r^2[\epsilon(r) - p(r)]}{g(r)} , 
\\
Q(r)&=&\frac{4\pi}{g(r)}\left[5\epsilon(r) + 9p(r) + 
\frac{\epsilon(r)+p(r)}{v_s^2(r)}- \frac{6}{4\pi r^2}\right]
\nonumber\\ 
&-& 4\left[ \frac{m(r)+4\pi r^3 p(r)}{r^2g(r)} \right]^2,
\label{qr}
\end{eqnarray}
with $v_s^2(r)=\partial p(r)/\partial\epsilon(r)$ being the squared sound velocity~\citep{tidal1,tidal2,tidal3,tidal4}. We show the results concerning $\Lambda$ in Fig.~\ref{lambda}.
From this figure, one notices that the inclusion of SRC in the system favors the model to attain the constraint of $\Lambda_{1.4}=190^{+390}_{-120}$~\citep{Abbott_2018} for parametrizations with $K_0=(240\pm 20)$~MeV. For the model presented here, it is also clear that the inclusion of SRC systematically decreases $\Lambda$ in all cases. The physical reason for this effect comes from the fact that SRC soften the EoS, as already discussed. In this case, the NS radius is also reduced by these correlations, and due to the relation given by $\Lambda\sim R^\alpha$, verified in different hadronic models for a 1.4 $M_\odot$ star for instance~\citep{defrmf,defsky}, it is straightforward to conclude that $\Lambda$ decreases with the radius decreasing. For the \mbox{CCS} model, this decrease makes the model compatible with the astrophysical data analyzed. Finally, we plot in Fig.~\ref{l1l2} the tidal deformabilities $\Lambda_1$ and $\Lambda_2$ of the binary neutron stars system with component masses $m_1$ and $m_2$ ($m_1>m_2$), related to the GW170817 event, and taking into account the range for $m_1$ given by $1.365\leqslant m_1/M_\odot \leqslant 1.60$~\citep{Abbott_2017}. The mass of the companion star $m_2$, is obtained from the relationship between $m_1$, $m_2$, and the chirp mass, that reads
\begin{align}
\mathcal{M}_c = \frac{(m_1m_2)^{3/5}}{(m_1+m_2)^{1/5}},
\end{align}
and is fixed at the observed value of $1.188M_\odot$, according to~\cite{Abbott_2017}. Upper and lower orange dashed lines correspond to the 90\% and 50\% confidence limits, respectively, provided by LIGO and Virgo Collaboration~\citep{Abbott_2018}. 
It is clear that the effect of SRC is to move the $\Lambda_1\times\Lambda_2$ curves of our excluded volume model to the region of compatibility with the LIGO and Virgo Collaboration data regarding the GW170817 event, due to the fact that SRC decreases the values of both dimensionless tidal deformabilities. In the figure, we also furnish a band with results obtained through the relativistic mean field models studied in~\citep{defrmf} that are consistent with constraints from nuclear matter, pure neutron matter, symmetry energy, and its derivatives analyzed in~\citep{dutra2014}. Notice that the parametrizations of the \mbox{CCS-SRC} model also have a good intersection with this band. 

It is also worth to noting that the \mbox{CCS-SRC} parametrizations used to construct Figs.~\ref{mrdata}, \ref{lambda}, and~\ref{l1l2} have the symmetry energy slope at the saturation density around $108$~MeV. This value is inside the range of $L_0=(106\pm 37)$~MeV, claimed in~\cite{piekarewicz} to be in full agreement with the updated results provided by the lead radius experiment (PREX-2) collaboration concerning the neutron skin thickness of $^{208}\rm Pb$~\citep{prex2}. Nevertheless, it is also important to mention that there are other studies pointing out smaller ranges for $L_0$. In~\cite{reinhard}, for instance, the interval of $L_0=(54\pm 8)$~MeV was determined from an analysis that takes into account theoretical uncertainties of the parity-violating asymmetry in $^{208}\rm Pb$. Ab initio calculations performed in~\cite{abinitio}, also for the $^{208}\rm Pb$ nucleus, predict the range of $L_0=(37-66)$~MeV for the slope parameter. Furthermore, according to~\cite{lattimer-slope}, the range of $L_0=(-5\pm 40)$~MeV is related to the results of the neutron skin thickness of $^{48}\rm Ca$ provided by CREX Collaboration~\citep{crex}. Another analysis in~\cite{zhang2022bayesian} combined the results from PREX-2 and CREX and found $L_0=15.3^{+46.8}_{-41.5 }$~MeV through a Bayesian inference. However, another combination of the PREX-2 and CREX results produced, through a covariance analysis, higher values for this isovector quantity: $L_0=(82.32\pm22.93)$~MeV~\citep{kumar}. We verified that for lower values of $L_0$, the \mbox{CCS-SRC} parametrizations are not simultaneously compatible with all astrophysical constraints depicted in Fig.~\ref{mrdata}. Moreover, in this case, the model produces extremely low values of $J$, for example, $J\sim 19$~MeV for $L_0=66$~MeV. This feature leads the bulk parameter space of the model with SRC to the direction of higher values of $L_0$. A more complete description, namely, the one in which lower values of $L_0$ are also allowed, necessarily imposes a suitable modification in the isovector sector. This specific study is already been performed for the \mbox{CCS-SRC} model.

\section{Summary and concluding remarks} 
\label{summ}

In this work, we have included, in a phenomenological way, SRC~\citep{baoli2015,baoli2016,cai2016} in a vdW-type model applied to the description of asymmetric nuclear matter. Excluded volume (EV) models have been recently used in relativistic hadronic systems~\citep{vovchenko2015a,vovchenko2015b,vovchenko2017a,vovchenko2017b,sagun}. It is an attempt to treat nuclear matter systems more realistically since it considers the nucleon as a finite spatial dimension object and no longer a structureless particle.
In~\citep{lourenco2019,jpgnosso}, in particular, authors developed a density-dependent vdW model in which attractive and repulsive parts of the nucleon-nucleon interaction were assumed depending on the nuclear density. Here we followed the same procedure and have used the Carnahan–Starling method for modeling the latter (see Eq.~\eqref{brho}), and a suitable expression for the former (see Eq.~\eqref{arho-ccs}) that ensures the structure of the Clausius real gas model~\citep{vovchenko2017a,vovchenko-clausius}. After the implementation of SRC in this theoretical framework, resulting in a model named $\mbox{CCS-SRC}$~(Clausius-Carnahan–Starling-SRC) model, we investigated its capability in correctly describing some features of both, nuclear and stellar matter. The four free parameters of the model are adjusted in order to reproduce saturation density ($\rho_0$), the binding energy of infinite nuclear matter, incompressibility, and symmetry energy (or symmetry energy slope, equivalently), with all these quantities evaluated at $\rho=\rho_0$.  

We verified that one of the effects of including SRC in the model is the shift of the break of causality to a higher-density region. It is important to mention that EV relativistic models suffer from this issue, namely, the break of causality due to the lack of a complete treatment of the Lorentz contraction for the finite-size nucleons. As we have shown, SRC helps to circumvent this problem in an effective way. We also observed that SRC did not destroy the linear relationship between symmetry energy and its slope ($L_0$), a correlation often found in the literature~\citep{jl1,jl2,jl3}. Furthermore, SRC increase the value of $L_0$ in comparison with the model without this phenomenology implemented. At higher density regime, another important finding shown in Fig.~\ref{press-high} is that the $\mbox{CCS-SRC}$ model completely satisfies the flow constraint, a wide constraint used to validate and select hadronic models~\citep{dutra2014}, for parametrizations constructed by running $K_0$ in the range of $K_0=(240\pm 20)$~MeV~\citep{garg2018}. 

With regard to the stellar matter, the inclusion of SRC in the CCS model softens the EoS generated as input to the TOV equations used to construct the mass-radius profiles. The opposite effect is observed in RMF models presenting quartic self-interaction in the repulsive vector field, namely, models in which the Lagrangian density presents a term given by $C_\omega(\omega^\mu\omega_\mu)^2$. For these models, SRC make the EoS stiffer and consequently capable of producing more massive neutron stars. However, RMF models in which $C_\omega=0$ exhibit the same behavior as the one found here, i.e., softer EoS in comparison with the ones without SRC added. Nevertheless, the \mbox{CCS-SRC} model still generates mass-radius diagrams compatible with recent astrophysical constraints, such as those coming from gravitational waves data related to the GW170817~\citep{Abbott_2017,Abbott_2018} and GW190425~\citep{Abbott_2020-2} events, data from the NICER mission regarding the pulsars PSR~J0030+0451~\citep{Riley_2019,Miller_2019} and PSR~J0740+6620~\citep{Riley_2021,Miller_2021}; and data from the latter pulsar extracted from~\cite{Fonseca_2021}. Our results show that SRC also favor the model to be consistent with the constraints regarding the dimensionless tidal deformability, namely, the one related to the $1.4M_\odot$, namely, $\Lambda_{1.4}=190^{+390}_{-120}$~\citep{Abbott_2018}, and those from the binary neutron stars system~\citep{Abbott_2017}, both of them provided by the LIGO and Virgo Collaboration through the analysis of gravitational waves detected in the GW170817 event. In this particular case, it was observed that SRC decrease the value of $\Lambda$ due to the reduction of the neutron star radius caused by the softening of the EoS. 

Finally, the values found for $L_0$ are inside the range of $L_0=(106\pm 37)$~MeV, pointed out in~\cite{piekarewicz} as compatible with data from the PREX-2 collaboration with regard to the $^{208}\rm Pb$ neutron skin thickness~\citep{prex2}. We also mention that, for the case in which lower values of $L_0$ are considered, the model is not able to simultaneously reconcile with all astrophysical constraints. Furthermore, very low values of $J$ are also found in this case. This feature has motivated us to investigate a possible improvement in the isovector sector of the model in order to make it suitable to also reach this particular region of the parameter space.

\section*{Acknowledgments}
This work is a part of the project INCT-FNA Proc. No. 464898/2014-5. E.~H.~R. is supported with a doctorate scholarship by Coordena\c c\~ao de Aperfei\c coamento de Pessoal de N\'ivel Superior (CAPES). This work is also supported by Conselho Nacional de Desenvolvimento Cient\'ifico e Tecnol\'ogico (CNPq) under Grants No. 312410/2020-4 (O.L.) and No. 308528/2021-2 (M.D.). O.L. and M.D. also acknowledge Funda\c{c}\~ao de Amparo \`a Pesquisa do Estado de S\~ao Paulo (FAPESP) under Thematic Project 2017/05660-0. O.~L. is also supported by FAPESP under Grant No. 2022/03575-3 (BPE). This study is also financed by CAPES – Finance Code 001 - Project number 88887.687718/2022-00 (M.~D.).

\section*{Data availability statement}
This manuscript has no associated data or the data will not be deposited. All data generated during this study are contained in this published article.

\bibliography{references-revised2}
\bibliographystyle{mnras}

\end{document}